\documentclass[journal,comsoc]{IEEEtran}
\usepackage{mathtools}
\usepackage{float}
\usepackage{graphicx} %package to manage images
\usepackage[colorinlistoftodos]{todonotes}
\graphicspath{ {images/} }

\usepackage[linesnumbered,ruled,vlined]{algorithm2e} % ,noend %remove "vlined" for puting "end"

\usepackage[utf8]{inputenc}
\usepackage[english]{babel}
\usepackage{amsthm,amssymb}

\usepackage{optidef}

\theoremstyle{definition}

\makeatletter
\renewcommand{\fnum@figure}{Fig. \thefigure}
\makeatother

\DeclareMathOperator{\sgn}{sgn}

\IEEEoverridecommandlockouts                              % This command is only
\title{%\LARGE 
Shifted Pruning for List Decoding of Polar Codes
}

\author{
\IEEEauthorblockN{Mohammad Rowshan, \textit{Student Member, IEEE} and Emanuele Viterbo, \textit{Fellow, IEEE}}
 \thanks{Mohammad Rowshan and Emanuele Viterbo are with the  department of electrical and computer systems engineering (ECSE), Monash University, Melbourne, VIC 3800, Australia. (e-mail: \{mohammad.rowshan, emanuele.viterbo\}@monash.edu).}
 \thanks{* This research work is supported by the Australian Research Council under Discovery Project ARC DP160100528.}\\
}

\begin{document}

\maketitle
\thispagestyle{empty}
\pagestyle{empty}

%%%%%%%%%%%%%%%%%%%%%%%%%%%%%%%%%%%%%%%%%%%%%%%%%%%%%%%%%%%%%%%%%%%%%%%%%%%%%%%%
\begin{abstract}
In successive cancellation list (SCL) decoding, the tree pruning operation retains the $L$ best paths with respect to metric at every decoding step. However, the correct path might be among the $L$ worst paths due to imposed penalties. In this case, the correct path is pruned and the decoding process fails. 
In this work, we propose a scheme for additional decoding attempts when decoding fails, in which the {\em pruning window} does not necessarily select the $L$ best paths, but this window is shifted between positions 1 and  $2L$  in the sorted list. In the simplest form, the $L$ worst paths are selected at the decoding step where the probability of elimination of the correct paths is high. 
Additionally, we generalize the scheme and propose a number of variants such as {\em constrained shifting, nested shifting} and {\em shifting under segmented decoding}, aiming to reduce the computational complexity. 
The numerical results for polar codes of length 512 with code rates 0.5 and 0.8 and list sizes $L=2, 8, 32$, show that the shifted-pruning scheme can provide 0.25-0.5 dB gain in error correction performance, while the average computational complexity approaches  the conventional list decoding complexity at practical FER ranges.
\end{abstract}

%%%%%%%%%%%%%%%%%%%%%%%%%%%%%%%%%%%%%%%%%%%%%%%%%%%%%%%%%%%%%%%%%%%%%%%%%%%%%%%%

%% The paper must be self-contained. However, if you are referring to
%% a full version for checking certain proofs, please provide the
%% publically accessible location below.  If the paper is completely
%% self-contained, you can remove the following line from your submission.
%\textit{A full version of this paper is accessible at: \url{http://www.isit2018.org/}}

\section{Introduction}
Polar codes proposed by Ar\i kan in \cite{arikan} are the first class of constructive channel codes that was proven to achieve the symmetric (Shannon) capacity of a binary-input discrete memoryless channel (BI-DMC) using a low-complexity successive cancellation (SC) decoder. %to the point that they have been chosen as a coding scheme for control channel in the 5th generation of mobile broadband standard \cite{3GPP}. 
However, error correction performance of polar codes under SC decoding is far form optimal. To address this issue,  successive cancellation list (SCL) decoder was proposed in \cite{tal} and yields an error correction performance comparable to maximum-likelihood (ML) decoding at high SNR. Furthermore, concatenation of cyclic redundancy check (CRC) bits to polar codes provide a significant improvement in the performance.

When SC or SCL decoding fails, we may be able to correct the error(s) in additional decoding attempts. 
Bit-flipping \cite{afisiadis} is a popular method to improve the error correction performance of SC decoder by flipping the low-reliability bit(s) when the decoding fails. As the experiments in \cite{afisiadis} showed, a predominant portion of the decoding failures occurs by a single error in bit estimation due to channel noise. Thus, by finding the first erroneous bit and flipping the estimated value, the error propagation can be prevented. This idea was further improved by using different schemes and metrics in selecting the low-reliability bits and multiple bit-flipping algorithms \cite{chandesris,zhang}. The performance of these methods can approach the performance of SCL decoding with average list size, which is inferior to the performance of CRC-aided SCL decoding. However, in terms of complexity, nested/multiple bit-flipping may require a massive number of attempts. %the size of index set in this methods grows exponentially by increasing the number of low-reliability bits, which results in a massive number of attempts.

Unlike the previous work, where the focus was improving the performance of SC decoding, other literature aimed at improving the the performance of SC list decoding. In \cite{yongrun}, the bit-flipping method was employed in CRC-aided SCL decoding in which the error correction performance improved by 0.15-0.25 dB for block-lengths of 256 and 512 and code rate $1/2$. 
Adaptive CRC-aided SCL decoding in \cite{li} improved the performance by gradually increasing the list size ($L$) by factor of two when the decoding fails. However, increasing the list size not only contributes in a larger complexity but also requires enough hardware resources to support that. Also, as the results show, the performance gain beyond list size $L>32$ is less then 0.1 dB which is obtained at a very high cost of doubling the resources. 
Lastly, in the repetition-assisted scheme for CRC-aided SCL decoding  \cite{rowshan2}, the low-reliability bits were repeated in the code to replace the erroneous bits in additional decoding attempts. This scheme improved the performance by 0.2 dB at high code rates.

The aforementioned methods either cannot outperform CRC-aided SCL decoding with large list size or require significantly larger hardware resources. A low-complexity decoding scheme which improves the performance significantly without the need to extra hardware resources is desirable.

In this work, instead of flipping the low-reliability bits to avoid error propagation in CRC-aided SCL decoding, we suggest to change the pruning rule in order to avoid the elimination of the correct path. This scheme can avoid multiple errors in bit-estimation (represented by accumulated penalties) due to the properties of the list decoding process. Thus, it can outperform the bit-flipping method \cite{yongrun} with a significantly lower number of attempts, resulting in lower complexity. Similar to the bit-flipping scheme, when the decoding fails, additional decoding attempts are performed in this scheme.

In summary, the main contributions of this work are as follows:

\begin{itemize}
    \item A scheme called shifted-pruning (SP) is proposed for list decoding in which the elimination of the correct path is prevented in the additional decoding attempts,
    \item The shifted-pruning scheme is generalized for any $k$-shift of pruning-window. This flexibility results in the reduction of the computational complexity through reduced number of decoding attempts,
    \item Nested shifting is proposed for further improvement of the error correction performance. Shifted-pruning under segmented list decoding is suggested as a simple realization of nested shifting which results in a significantly lower computational complexity.
\end{itemize}

\textbf{Paper Outline:}
The rest of the paper is organized as follows. Section II introduces the notation for the polar codes and describes the SC and SCL decoding. Section III studies the event of the elimination of the correct path. In Section IV, the solution for avoiding the elimination of the correct path is proposed. Sections V-VIII cover the generalization and various implementations of the proposed scheme. In Section IX, the implementation results are shown. Finally Section X makes concluding remarks.

\section{Preliminaries}

%\subsection{Polar Codes}
A polar code of length $N = 2^n$ with $K$ information bits is denoted by $P(N,K,\mathcal{A})$, where $\mathcal{A}$ is the index set of the non-frozen bits. The polar codes' generator matrix is $G_N = G_2^{\otimes n}$, where
$G_2 \overset{\Delta}{=}{\footnotesize \begin{bmatrix}
1 & 0 \\
1 & 1
\end{bmatrix} }$, and $(\cdot)^{\otimes n}$ denotes the $n$-th Kronecker power \cite{arikan}. Polar codes are encoded by $x_0^{N-1}\!=\!u_0^{N-1}G_N$ where $u^{N-1}_0 = (u_0, u_1, . . . , u_{N-1} )$ is the input vector, including frozen and non-frozen bits, and $x^{N-1}_0\!=\!(x_0, x_1, . . . , x_{N-1} )$ represents the coded bits vector. Let $y^{N-1}_0\!=\!(y_0, y_1, . . . , y_{N-1} )$ denote the output vector of a noisy channel.

%Suppose $W:\mathcal{X \to Y}$ denote a binary discrete memoryless channel with channel transition probabilities $\{W(y|x) : x \in \mathcal{X}, y \in \mathcal{Y}\}$ where $\mathcal{X}$ and $\mathcal{Y}$ are input and output alphabets, respectively. After channel polarization, the transition probability of the i-th subchannel is given by

%\begin{equation}
%\label{eq:trans_prob}
%W^{(i)}_N(y_1^N, u_1^{i-1}|u_i ) = \sum_{\substack{u^N_{i+1} \in \mathcal{X}^{N-i}}}
%\frac{1}{2^{N-1}} W_N (y^N_1|u^N_1),
%\end{equation}

%where $W_N(y^N_1|u^N_1) = \prod_{i=1}^N W(y_i|x_i)$ is the transition probability between $u^N_1$ and $y^N_1$.

%The information bits are transmitted through the $K$ most reliable subchannels while the remaining sub-channels are used for sending the frozen bits, usually set to zero.

%\subsection{Successive Cancellation (SC) Decoding}
%SC decoding algorithm is based on the greedy tree search. 
Let $\lambda^i_j$ denote the log-likelihood ratio (LLR) of  bit $i$ at stage $j$ of factor graph of SC decoding. The non-frozen bits are estimated successively based on the decision LLR, $\lambda^i_0$,  via a one-time-pass through the factor graph. %in Fig. \ref{fig:factor_graph}. 
Successive hard decisions make the SC solution sub-optimal. When decoding the $i$-th bit, if $i \notin \mathcal{A}$, 
$\hat{u}_i=0$, as $u_i$ is a frozen bit. Otherwise, bit $u_i$ is decided by a maximum likelihood (ML) rule %or binary quantizer function
$h(\lambda^i_0)$ in (\ref{eq:sc_hard_decision}), which depends on the estimation of previous bits, i.e. $\hat{u}_0, ..., \hat{u}_{i-1}$.
\begin{equation}
\label{eq:sc_hard_decision}
\small
\hat{u}_i = h(\lambda^i_0)= \begin{dcases*}
        0 & $\lambda^i_0 = \ln \frac{P(Y,\hat{u}_0^{i-1}|\hat{u}_i=0)}{P(Y,\hat{u}_0^{i-1}|\hat{u}_i=1)}>0$,\\
        1 & otherwise\\
\end{dcases*}
\end{equation}

% \begin{figure}
%     \centering
%     \includegraphics[width=0.8\columnwidth]{factor_graph_4b_sb}
%     \caption{\vspace{-5mm}The factor graph of SC decoder for $N=4$} %\cite{rowshan}}
%     \label{fig:factor_graph}
%     \vspace{-5pt}
% \end{figure}
\normalsize

%In order to estimate the transmitted, the decoder computes the internal LLR of the indexed edges  as follows:
%\small
%\begin{equation}
%\label{eq:edge_llr}
%\lambda_i^j = \begin{dcases*}
%        f(\lambda^i_{j+1},\lambda^{i+2^j}_{j+1})  & if $B(i,j)=0$\\
%        g(\lambda^{i-2^j}_{j+1},\lambda^i_{j+1},\hat{\beta}^{i-2^j}_j)  & if $ B(i,j)=1$\\
%\end{dcases*}
%\end{equation}
%\normalsize

%where $0\!\leq\!i<\!N$, $0\!\leq\!j\!\leq\!n$, $B(i,j)\!=\!\lfloor \frac{i}{2^j}\rfloor\mod 2$ and $\hat{\beta}_i^j$ denotes the partial sum of bits, which corresponds to the propagation of estimated bits $\hat{u}_i$ back into the factor graph.

%The $f$ and $g$ functions are well approximated by \cite{balatsoukas}:
%\begin{equation}
%\label{eq:f_func}
% f(\lambda_a,\lambda_b)=\sgn(\lambda_a)\cdot \sgn(\lambda_b)\cdot \min(|\lambda_a|,|\lambda_b|)
%\end{equation}
%\begin{equation}
%\label{eq:g_func}
% g(\lambda_a,\lambda_b,\hat{\beta})=(-1)^{\hat{\beta}}\lambda_a+\lambda_b
%\end{equation}
%where $\lambda_a$ and $\lambda_b$ are incoming LLRs to the node and $\hat{\beta}$ is a partial sum of previously decided bits.

%\subsection{Successive Cancellation List (SCL) Decoding}
Unlike SC decoding which makes a final decision at each step, SCL decoding considers both possible values $u_i= 0$ and $u_i=1$. In SCL decoding, the $L$ most reliable paths are preserved at each decoding step to avoid growing the number of paths exponentially. Let $\hat{u}_i[l]$ denote the estimate of $u_i$ in the $l$-th path, where $l\!\in\!\{1, 2, \ldots ,L\}$. In \cite{balatsoukas} unlike \cite{tal}, a path metric (PM) based on LLRs magnitude is used to measure the reliability of the paths. PM at $\hat{u}_i[l]$ is approximated by:
\begin{equation}
\label{eq:pm_func} \small
PM^{(i)}_l\! = \!
\begin{dcases*}
PM^{(i-1)}_l + |\lambda^i_0[l]| & if $\hat{u}_i[l]\!\neq\!\frac{1}{2} (1\!-\!\sgn(\lambda^i_0[l]))$ \\
PM^{(i-1)}_l & otherwise	\\
\end{dcases*}
\end{equation}
\normalsize
where $PM^{(-1)}_l = 0$.

As (\ref{eq:pm_func}) shows, the path of the less likely bit value is penalized by $\lambda^i_0$ of that bit. The $L$ paths with smallest path metrics are chosen from $2L$ paths at the $i$-th step and stored in ascending order from $PM^{(i)}_1$ to $PM^{(i)}_L$.
After decoding the $N$-th bit, the path with the smallest path metric $PM^{(N)}_1$ is selected as the estimated codeword.

%\subsection{CRC-aided SCL Decoding}
Additionally, when the SCL decoding fails, the correct path might still be in the list but not in the position of the most likely path. Adding an $r$-bit CRC as an outer code to the information bits can assist the decoder in error detection and finding the correct path among the $L$ paths. However, this concatenation increases the polar code rate to $(K+r)/N$. In this paper, $P(N,K+r)$ denotes a polar code of length N with K information bits concatenated with $r$-bit CRC.

\section{Elimination of Correct Path}\label{sec:elim}
The path metric in LLR-based SCL decoding is in fact  the accumulated penalties over the bit-channels that do not satisfy the condition in (\ref{eq:pm_func}). The conventional SCL decoding algorithm retains $L$ paths with smallest $PM$ values, i.e. the paths that are penalized less. If the correct path is penalized multiple times or penalized once with a large $|\lambda_0^i|$, it may be pruned from the list and the decoding fails. In this section, we analyze the elimination events numerically, and we explain in detail the elimination event.

\subsection{Numerical Analysis} \label{ssec:num_analysis}
\begin{figure}
    \centering
    \includegraphics[width=1\columnwidth]{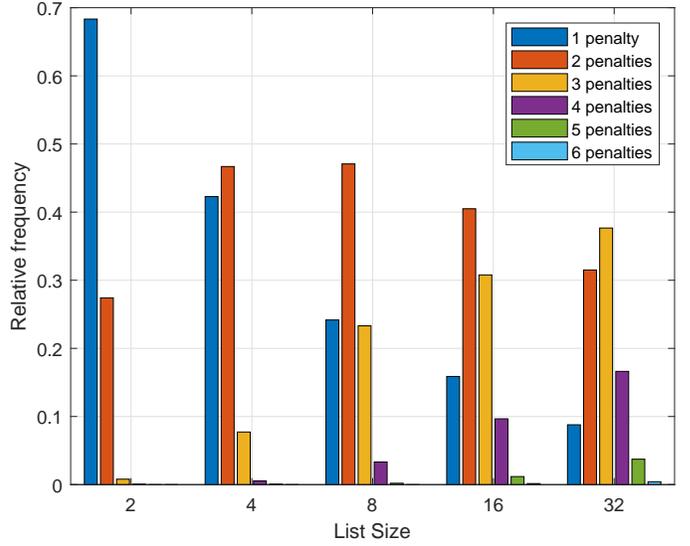}
    \caption{Relative frequency of number of penalties leading to elimination of the correct path at different list sizes for $N=256$ and $R=0.5$}
    \label{fig:pen_distrib}
    \vspace{-5pt}
\end{figure}

Fig. \ref{fig:pen_distrib} illustrates the number of penalties ($p=1,2,...,6$) causing the elimination of the correct path in the CRC-aided SCL decoding with 8-bit CRC generator polynomial $g(x)=x^{8}\!+\!x^{7}\!+\!x^6+\!x^4+\!x^2\!+\!1$ at different list sizes. These data are collected over 2000 block error occurrences at FER=$10^{-3}$. This figure shows that as the list size increases, a larger number of penalties is required to eliminate the correct path. In other words, at a large list size, the correct path tolerates more number of penalties and the elimination of the correct path caused by small number of penalties reduces. This is the reason for error correction gain at large list sizes.

Another interesting observation in Fig. \ref{fig:pen_distrib} is that a single penalty is the cause of less than 20\% of the block errors at a relatively large list size (e.g. $L=16, 32$). Our observation shows that these single penalties occur over the bit-channels with relatively large $|\lambda_0^i|$ where $PM_i$ of the correct path because the penalty becomes larger than the {\em path metric range} \cite{rowshan}, \mbox{\em PMR}$_i=PM^{(i)}_L-PM^{(i)}_1$, and consequently the correct path is pruned. It was shown in \cite{rowshan} that \mbox{\em PMR} value is near or at local minimum over these bit-channels.

\begin{figure}
    \centering
    \includegraphics[width=1\columnwidth]{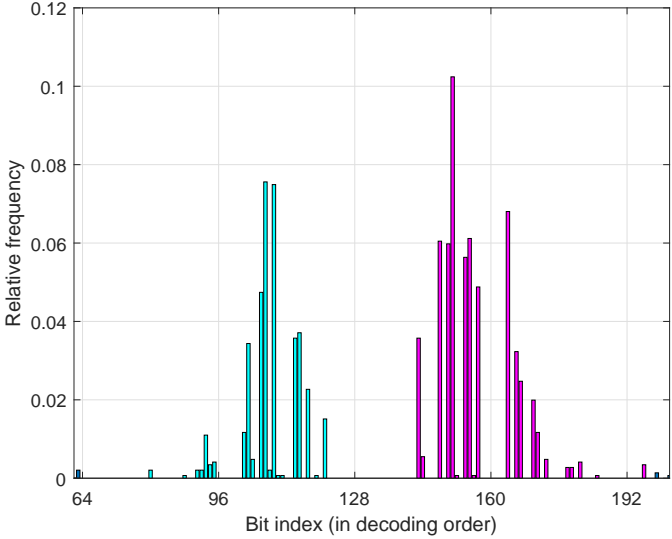}
    \caption{Relative frequency of elimination caused by more than one penalty over bit-channels for $N=256$, $R=0.5$ and $L=8$}
    \label{fig:elim_bit}
    \vspace{-5pt}
\end{figure}

Fig. \ref{fig:elim_bit} shows the bit positions where the event of elimination of the correct path due to occurrence of penalties. This set of bit indices are called {\em vulnerable bits}, $\mathcal{S}$. The relative frequencies of eliminations in two middle sub-blocks of $N=256$ are shown by different colors. One can observe a trend of increasing the frequency of elimination from the beginning of sub-block, reaching a peak and then decreasing the frequency. We will explain the reason behind this trend based on the error probability of bit-channels in the next section, and then we will define a set of bits that contribute the most in the penalties.

\subsection{Critical Set}
The key to the proposed scheme is to know the indices of the bits which contribute the most in the elimination of the correct path. There exist some heuristic methods in the literature that are used for identifying the first erroneous bit in the SC-flip scheme. The bits which are prone to the occurrence of the first error are equivalent to the first penalized bit of the correct path. Hence, these bits also contribute in the elimination of the correct path due to one or more penalties.
The index set of these bits can be determined offline in a channel-independent manner which is called {\em critical set} \cite{zhang}. In \cite{afisiadis}, the bits with smaller amplitude of log likelihood ratio (LLR) in the previous decoding attempt are assumed the bits which are more prone to error in SC decoding.  
As another channel-dependent method, the numerical results for decoding $P(256,128+8)$ with $L=8$ and 8-bit CRC at $E_b/N_0=2.5$ dB show that the non-frozen bit-channels %with the minimum row-weight in the sub-blocks,  
which experience a \mbox{\em PMR} drop (i.e. \mbox{\em PMR$_i$-PMR$_{i-1}<0$}), account for about 99.68\% of the positions that the penalty occurs over them.  This dynamic set of bits collected from a several iterations is equivalent to the static {\em critical set} \cite{zhang}, denoted by $\mathcal{CS}$, used in SC bit-flip decoding. %Algorithm \ref{alg:cs} illustrates the procedure of finding the elements of $\mathcal{CS}$. %These methods provide an index set of bit-channels with high probability of penalty occurrence but these sets are not ordered based on the frequency of elimination. The easiest way to find an ordered index set is using Monte Carlo experiment. Note that the  bit-channels with the lowest reliability are not the bits with the highest occurrence of elimination of the correct path.

\subsection{How the correct path is eliminated?}
Let us consider a segment of bit-channels in which a sufficient number of high-reliability frozen bits (reliability measure is obtained from one of the code construction methods such as approximated average for decision LLR, $\lambda_0^i$ in \cite{trifonov}) exist in the beginning of the segment as an independent segment because the previous segments do not have a significant effect on the elimination events in the current segment. This relative independency is originated from the {\em recovery phenomenon} in which the penalized correct path moves towards the position in the list with the highest likelihood (position 1 in the sorted list) \cite{rowshan3}, i.e. in the presence of enough high-reliability frozen bits, the effect of penalties is neutralized to some extent.
In polar codes of length $N=256$ and $R=1/2$, these independent segments are sub-blocks of length $N_{sb}=64$. 
%As our observation shows, the relative frequency of elimination over the bit-channels with row-weight 16 in $G_N$ located in the 2nd sub-block (bit indices $65-128$) is significantly larger. In Fig. \ref{fig:elim_bit}, they are shown with bars of height at least 0.01. Thus, the bit channels with minimum row-weight in the sub-block $j$ forms a set of {\em vulnerable bits} $\mathcal{V^j}$. 
%Now, if the proceeding sub-block have a local minimum row-weight of non-frozen bits equal to previous sub-block, we can merge the two adjacent sub-blocks and form a larger set of vulnerable bits.

Let $P_i$ and $E_i$ denote the random variables indicating the events that the correct path is or is not penalized and eliminated by values 1 and 0, respectively, over bit $i$. 

\begin{figure}
    \centering
    \includegraphics[width=1\columnwidth]{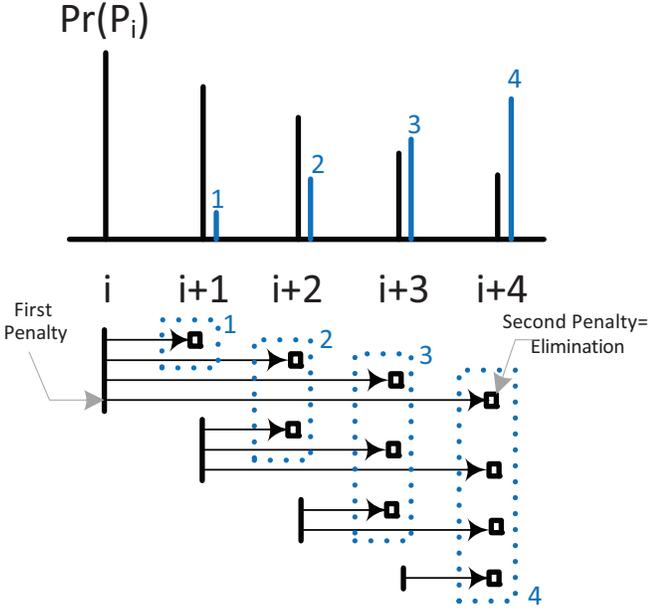}
    \caption{All the combinations resulting in two accumulated penalties for bit $i+1$ to bit $i+4$}
    \label{fig:2-pen_contrib}
    \vspace{-5pt}
\end{figure}

Elimination of the correct path occurs when its metric $PM_{l_c} > PM_L$. This is the result of (accumulated) penalties imposed on the correct path according to (\ref{eq:pm_func}). In section \ref{ssec:num_analysis}, we showed that when $L>2$, most of the elimination events occurs due to more than one penalty. Fig. \ref{fig:2-pen_contrib} illustrates how possibility of accumulating only two penalties from bit $i+1$ to $i+4$ increases by factor of $\binom{b}{1}$ where $b=1,2,...,4$. Note that here, we assume the second penalty occurs on the current bit and the first penalty on one of the previous bits. For example, if the second penalty occurs on bit $i+1$, the first penalty could have only occurred on bit $i$, while in case of the occurrence of the second penalty on bit $i+4$, the first penalty could have occurred on bit $i$, $i+1$, $i+2$, or $i+3$. Now, let $p$ and $j$ denote the number of penalties and the index of current bit in the segment ($j=0,1,...$), respectively. In general, the probability that $p$ penalties occur on bit $j$ is $P_{j}^p=p_{e,j}\cdot \binom{j}{p-1}\cdot\bar{p}_{e}^{p-1}\cdot (1-\bar{p}_{e})^{j-(p-1)}$ where $p_{e,j}$ is the error probability of bit $j$ and $\bar{p}_{e}=\sum_{k=0}^{j-1} p_{e,j}$ over the bits in the critical set. As Fig. \ref{fig:2-pen_contrib} shows, $Pr(P_i)=p_{e,j}^{P_i}\cdot (1-p_{e,j})^{1-P_i}$ decreases in a segment (when $P_i=1$, i.e. the penalty is imposed) and so the average probability $\bar{p}_{e}$ while the factor $\binom{j}{p-1}$ increases. These opposite trends result in a peak for the probability that $p$ penalties occur. The reason why a peak appears can be explained by assuming two trends in the form of two linear equations. Then, by knowing that the resulted equation from multiplication of these two linear equations is quadratic, we can realize the existence of a peak somewhere in the middle of the domain. This resembles the quadratic-like shape of probability of elimination shown in Fig. \ref{fig:elim_bit}.   %Hence, this probability first increases and after reaching to the peak at some bit starts decreasing. 
Note that the probability of elimination  of the correct path differs from the probability of accumulated penalties. The penalties imposed on the correct path moves it within the list of candidate paths from index $l=1$ towards $l=L$. The more penalties are imposed, the larger movement in the list occurs. While the correct path may move in the list over the bits in the critical set, another phenomenon is happening. From \cite{rowshan}, we know that $\mbox{PMR}=PM_L-PM_1$ decreases over the bits in the critical set. Considering this phenomenon along with movement of the correct path in the list, one can realize that the probability of event $PM_{l_c} > PM_L$ increases by decreasing the value of $PM_L$ even without imposing any penalty. Thus, the peak of relative frequency of elimination in Fig. \ref{fig:elim_bit} differs from the peak for the probability that $p$ penalties occur.  Unfortunately, we cannot provide an accurate model to estimate the probability of elimination of the correct path but one can see the factors involved in the elimination. %Our observation shows that the inverse of the average \mbox{PMR} over random decoding  iterations can indicate approximately the relative elimination frequency. That is, the lower the $\overline{\mbox{PMR}}$ is, the higher the probability of elimination of correct path will be. To formalize the use of $\overline{\mbox{PMR}}$ as an approximate indicator of $Pr(E_i=1)$, one can use $\frac{\overline{\mbox{PMR}}_{max}-\overline{\mbox{PMR}}_j}{\overline{\mbox{PMR}}_{max}}$ over the bits exist in the critical set of the segment.

\section{Shifted-pruning over Critical Set}
The correct path is pruned from the list when it has a relatively large $PM$ due to the accumulated penalties and falls among the paths with indices $L+1$ to $2L$ in the sorted list. Thus, one can think of changing the rule for pruning over the critical set to avoid the elimination of the correct path.
\begin{figure}
    \centering
    \includegraphics[width=1\columnwidth]{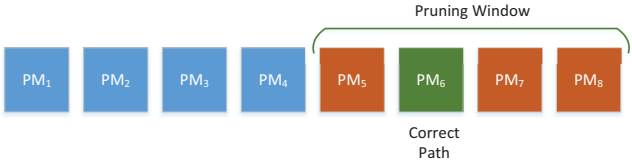}
    \caption{Shifting the pruning window by $k=L$ where $L=4$}
    \label{fig:pruning_window}
    \vspace{-5pt}
\end{figure}

\begin{figure}
    \centering
    \includegraphics[width=1\columnwidth]{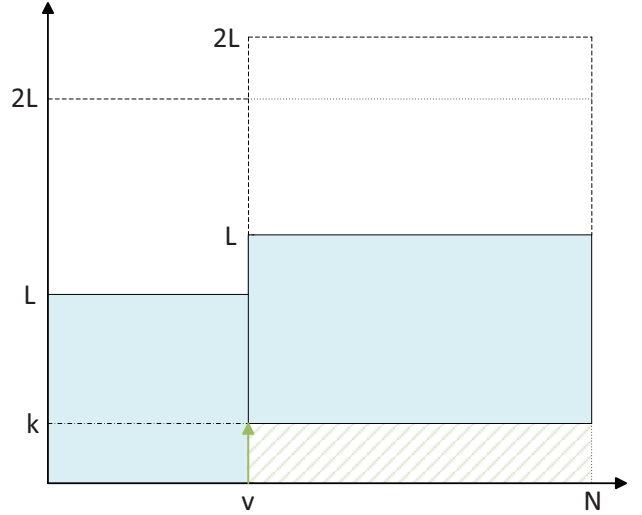}
    \caption{Shifting $k$ paths during list pruning operation at bit $v\in \mathcal{V}$}
    \label{fig:shifting}
    \vspace{-5pt}
\end{figure}

In the LLR-based SCL decoding, the most effective way to avoid pruning the correct path is to retain the $L$ paths with the largest $PM$s (instead of $L$ paths with the smallest $PM$ values) over the bits in the set $\mathcal{CS}$. This operation is named {\em shifted-pruning} because in the process of selection of the paths to remain in the list, the reference for the first path is shifted by $k$, i.e. we select the path $k+1$ to path $L+k$ (instead of path $1$ to path $L$) in the $PM$-based ordered set of paths. Fig. \ref{fig:shifting} shows shifting by $k$ paths over bit $v$. %However, there is a more efficient scheme to prune the decoding tree.

According to section \ref{ssec:num_analysis}, the elimination of the correct path in list decoding may occur due to multiple penalties, i.e. several errors in the estimation of the bit values due to the violation of the condition in (\ref{eq:pm_func}). Thus, single bit-flipping cannot avoid the elimination in many cases, particularly when a large list size is used. However, shifted-pruning can be effective because it tries to retain the correct path in the list regardless of number of penalties occurred. The numerical results in the next section show the effectiveness of the shifted-pruning scheme over the bit-flipping technique in avoiding multiple errors.

\begin{algorithm}
\label{alg:shifted-pruning}
\caption{List Decoding with Shifted-pruning}
%\SetAlgoLined
\DontPrintSemicolon
%\SetKwData{Left}{left}
%\SetKwData{This}{this}
%\SetKwData{Up}{up}
%\SetKwFunction{Union}{Union}
%\SetKwFunction{FindCompress}{FindCompress}
\SetKwInOut{Input}{input}
\SetKwInOut{Output}{output}
\SetKwRepeat{Repeat}{do}{while} %Defualt: \SetKwRepeat{Repeat}{repeat}{until}
\SetKwFunction{func}{Subroutine}
%\KwResult{Set of non-frozen bit indices, }
\Input{the received vector $y_1^N$, non-frozen set $\mathcal{A}$, $L$}
\Output{the recovered message bits $\hat{u}_1^N$}

    $t \gets 0$\;
    $k \gets L$\;
    crcPass $\gets$ {\em false}\;
    $\mathcal{CS}$ $\gets$ generateCS($\mathcal{A}$, $\log_2N$)\;
    \Repeat{$t \leq |\mathcal{CS}|$ \textsc{and} \textnormal{crcPass} = false}{ %\textsc small capitals
        $\hat{u}_1^N[1..L]$ $\gets$ SCLD($y_1^N$,$\mathcal{A}$, $\mathcal{CS}$, $L$, $t$, $k$)\;
        \For{$l\gets 1$ \KwTo $L$}{
            \If{\textnormal{CRC($\hat{u}_1^N[l]$)} = success}{
                $\hat{u}_1^N$ $\gets$ $\hat{u}_1^N[l]$\;
                crcPass = {\em true}\;
            }
        }
        \If{\textnormal{crcPass} = false}{
            $t\gets t + 1$\;  %// Freezing a vulnerable bit:\;
        }
    }
    \KwRet $\hat{u}_1^N$\;
    
    \SetKwFunction{FDecode}{SCLD}
    \SetKwProg{Fn}{subroutine}{:}{}
    \Fn{\FDecode{$y_1^N$,$\mathcal{A}$, $\mathcal{CS}$, $L$, $t$, $k$}}{
        $\mathcal{L} \gets \{1\}$ \tcp*{a single path in the list}
        \For{$i\gets 1$ \KwTo $N$}{
            Perform standard SCL Decoding:\;
            - Path pruning when $i\in\mathcal{A}$ AND $|\mathcal{L}| > L$:\;
            \uIf{$\mathcal{CS}(t) \neq i$}{
                $\mathcal{L} \gets \{1,...,L\}$ 
            }
            \Else{
                $\mathcal{L} \gets \{k+1,...,L+k\}$ 
            }
        }
        \KwRet $\hat{u}_1^N[1..L]$\;
    }
    \SetKwFunction{FDecode}{generateCS}
    \SetKwProg{Fn}{subroutine}{:}{}
    \Fn{\FDecode{$\mathcal{A}$, $n$}}{
        $B \gets \{-1\}$ \tcp*{initialize $B_{(n+1)\times 2^n}$}
        $cnt \gets 0$\;
        \For{$k\gets 1$ \KwTo $2^n$}{
            $B(n+1,k) \gets 0$\;
            \uIf{$k \notin \mathcal{A}$}{
                $B(n+1,k) \gets 1$ 
            }
        }
        \For{$i\gets n$ \KwTo $1$}{
            \For{$j\gets 1$ \KwTo $2^{i-1}$}{
                $B(i,j) \gets B(i+1,2j-1)+ B(i+1,2j)$\;
            }
        }
        \For{$i\gets 1$ \KwTo $n+1$}{   %\tcp*{tree levels}
            \For{$j\gets 1$ \KwTo $2^{i-1}$}{
            $x_1 = x_2 = j$\;
            \If{$B(i,j) = 0$}{
                \For{$k\gets 1$ \KwTo $(n+1)-i$}{
                    $x_1 = 2 x_1 - 1$\;
                    $x_2 = 2 x_2$\;
                    \For{$p\gets x_1$ \KwTo $x_2$}{
                        $B(i+k,p) \gets -1$
                    }
                    $\mathcal{CS}(cnt++)=x_1$
                }
            }
            }
        }
        \KwRet $\mathcal{CS}$\;
    }
\end{algorithm}
%\vspace{-15pt}

Algorithm \ref{alg:shifted-pruning} illustrates a modified list decoder to implement the shifted-pruning scheme. In this paper, the length of the shift is equal to the list size ($L$) as assigned in line 2. The critical set is generated using subroutine $generateCS$ %\todo{\tiny missing algorithm?}
which is called  in line 4. In the modified SC list decoding, if the decoding fails (line 11), the shifted-pruning scheme is performed on one of the bits in the set $\mathcal{CS}$ at every decoding attempt. The maximum number of attempts equals the size of the critical set, $|\mathcal{CS}|$ as shown in line 13. The shifted-pruning operation is performed in line 23 when decoding bit $i \in \mathcal{CS}$.

\section{Constrained Shifted-pruning Scheme}
Suppose the correct path is only eliminated at a bit index in the critical set. This assumption is close to reality as the performance of oracle-assisted and $CS$-based shown in Fig. \ref{fig:oracle-assisted} are almost the same. When the critical set elements are not sorted in order of probability of elimination, we may need to try all the bit indices in $\mathcal{CS}$ to apply shifted-pruning. %in order to avoid the correct path elimination. 
This imposes a large time complexity. For reducing the complexity, one or both of the following methods can be applied:
\subsection{Prioritizing the Bit-positions in Shifted-pruning}
The bit indices in $\mathcal{CS}$ can be sorted in descending order of probability of elimination of the correct path. Hence, in re-decoding iteration(s), first the shift is applied on the pruning operation of the bit index with the highest probability/frequency of elimination. This way, less number of iterations will be required to correct the received message and consequently the complexity reduces. Now, the question is how to find the priority of low-reliability bit-channels? As mentioned in Section \ref{sec:elim}, we do not have an exact model for the elimination occurrence. Monte Carlo method is an easy way to extract the relative frequency of elimination of correct path. %Another approximate method is using the average PMR by giving priority to the bit indices with smaller PMR. 
\subsection{Selecting a Subset of Critical Set}
Having the elements of $\mathcal{CS}$ sorted based on the priority for shifting, we can apply shifted-pruning on a subset of $\mathcal{CS}$ in which a majority of eliminations occur. In this method, reduction in time complexity is traded with a slight degradation in the error correction performance relative to use of full $\mathcal{CS}$. The amount of degradation is inversely proportional to the improvement in complexity.

\section{Generalized Shifted-pruning Scheme}
\label{sec:generalized SP}
The amount of shifting the pruning-window, $k$, plays an important role in the obtained complexity and error correction performance. Perhaps by $k=L$, the best possible performance can be obtained. However, when $k<L$, e.g. $k=L/2$, depending on the position of the correct path among $2L$ sorted paths, it can still avoid the elimination of the correct path. The interesting result of applying  $k<L$ shifts on the  bit index $i$ is avoiding the elimination at the bit index $i+1$ or a bit farther given that both indices are in $\mathcal{CS}$ and the correct path is in position $l<L$ where shifted pruning-window covers it.  Therefore, using a right subset of $\mathcal{CS}$ may not result in a significant degradation of performance while it may reduce the complexity significantly. You will see the complexity reduction obtained by $k=L/2$ over a small subset of critical set in Section \ref{sec:results}.

\begin{figure}
    \centering
    \includegraphics[width=1\columnwidth]{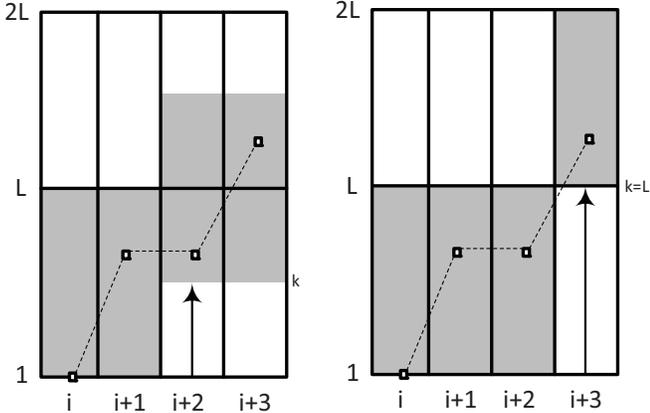}
    \caption{$k$-shift ($k<L$) vs $L$-shift during pruning operation.}
    \label{fig:k-shift}
    \vspace{-5pt}
\end{figure}

As a general scheme for shifting the pruning window, $k$ can vary in interval $0\leq k_i\leq L$ on different bit indices, $0\leq i\leq N-1$. Obviously, $k_i=0$ for $i\in \mathcal{A}^c$ and high-reliability bit indices. Finding a practical method to determine $k_i$ is an open problem. Studying the movement of the correct path  using Monte Carlo method may provide a set of patterns for $\{k_i\}$. A good pattern is the one that can reduce the complexity substantially at low degradation cost. 
In section \ref{sec:results}, we will show the results of some simple patterns for the variable shifting scheme. Additionally, employment of $k$-shift instead of $L$-shift can relax the problem of finding the exact position of the elimination. As Fig. \ref{fig:k-shift} shows, the shifting can be applied before the exact bit-position where the correct path is potentially pruned. The doted line in Fig. \ref{fig:k-shift} illustrates the movement of the correct path from one bit to another within the ordered list of paths.

\section{Nested Shifted-pruning Scheme}
\begin{figure}
    \centering
    \includegraphics[width=1\columnwidth]{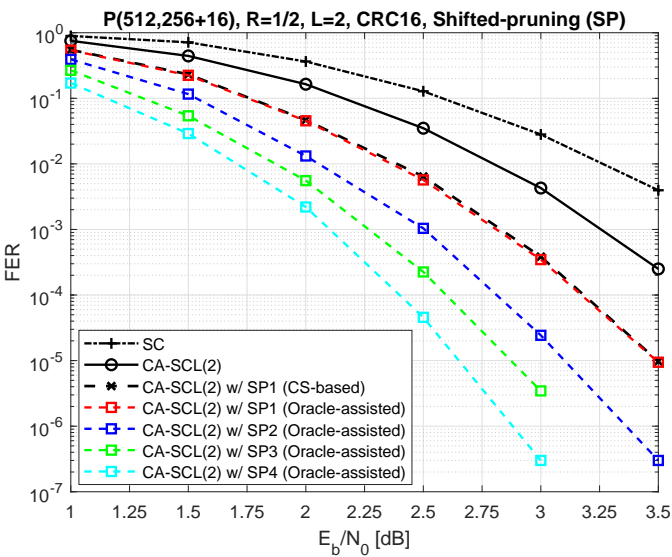}
    \caption{Performance of Multiple Shifts by an Oracle}
    \label{fig:oracle-assisted}
    \vspace{-5pt}
\end{figure}

\begin{figure}
    \centering
    \includegraphics[width=1\columnwidth]{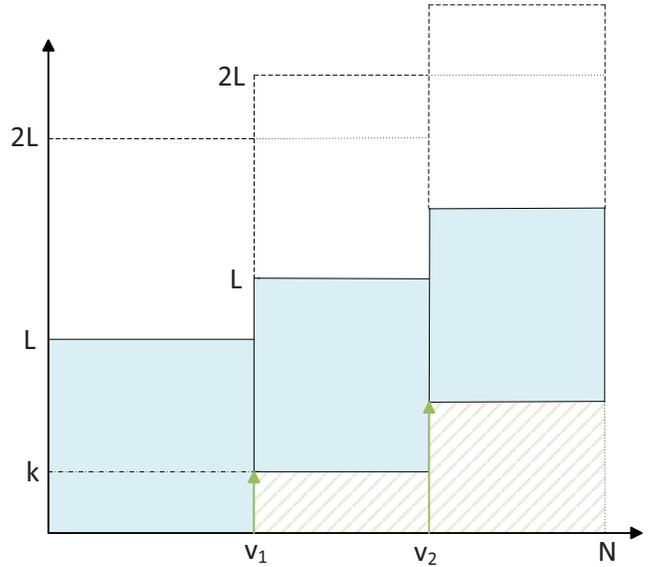}
    \caption{Nested shifting by $k$ paths at bit $v_1, v_2\in \mathcal{V}$ during pruning operation}
    \label{fig:shifting-nested}
    \vspace{-5pt}
\end{figure}

Elimination of the correct path may not be prevented by just one time shifting throughout decoding a codeword.  Suppose we have an oracle that can avoid the elimination at any bit indices, we can observe that there are cases that require more one-time shifting. Fig. \ref{fig:shifting-nested} illustrates an example of twice shifting. As Fig. \ref{fig:oracle-assisted} shows the error correction gain obtained from multiple shifts is significantly higher than one-time shifting. The notation "SP" in the figure is used for oracle-assisted shifted-pruning and x in "SPx" indicates the number of shifts applied throughout one decoding iteration to avoid the elimination of correct path. As can be seen in Fig. \ref{fig:oracle-assisted}, if we use the full critical set in additional decoding attempts, the performance is the same as the oracle-assisted performance. As the other dashed curves shows, employment of nested shifts could improve the performance further. However, finding the right combination of the bit indices for shifting requires a massive search. 

\section{Segmented Shifted-pruning}
\label{sec:seg SP}
To reduce the complexity of nested shifts, a heuristic metric might help. Here, we suggest to simplify the problem by the assumption that each elimination (when the elimination may happen  multiple times while avoiding the previous one(s)) occurs in one segment of the code-block. The observation shows that this assumption is close to reality given that the right length is chosen for the segments.  Unfortunately, in segmented list decoding \cite{guo}, we have to use multiple short CRCs. Note that the probability of undetected error by a short CRC is high \cite{rowshan2}. Considering that in the shifted-pruning scheme, we need to run additional decoding, consequently, the probability that an incorrect path is detected by CRC as the correct path increases significantly. Thus, the bad news is that we cannot expect to obtain the performance of nested shifts as shown in Fig. \ref{fig:oracle-assisted}. However, there is a good news which is a significant reduction in the computational complexity by using segmented list decoding. The reduction in complexity comes from the fact that we do not need to apply shifting on the whole code-block. For example, in decoding P(512,256+2*8) with two segments for which 8-bit CRCs are used, if the elimination occurs in the first segment only, we just apply the additional decoding attempts on that segment and once the elimination is avoided, the second segment does not require additional attempts for decoding it. Therefore, the additional computational complexity introduced by additional attempts halves for this codeword.

Note that the failures through detecting incorrect paths due to employing short CRCs are traded off by successes due to multiple shifts and overall performance improvement remains almost the same at low list size.

For computational complexity comparisons, since the block-length $N$ is fixed, we drop $N\cdot \log N$ from the computational complexity $O(LN\log N)$, hence, we use the average list size $L$ as a measure of complexity of the shifted-pruning scheme. Now, let denote the total iterations and total decoded messages during decoding, and number of segments by $t$, $c$, and $s$, respectively, the computational complexity of non-segmented and segmented decoding schemes are computed by $O(\frac{t}{c}\cdot L)$ and $O(\frac{t}{s\cdot c}\cdot L)$. Note that in the segmented decoding, $t$ refers to the iteration in each segment; hence the total iterations are sum of iterations at all segments.

\section{Numerical results}\label{sec:results}
To evaluate the performance of shifted-pruning scheme, Polar codes of length $N\!=\!2^{9}$ with the code rates of $R=K/N=0.5$ and $0.8$ are constructed using density evolution under Gaussian approximation \cite{trifonov} while optimized for high SNRs with design-SNRs of 5 dB and 4 dB, respectively. The LLR-based CRC-aided (CA) SCL decoder is used with 16-bit CRC generator polynomial of  $g(x)=x^{16}\!+\!x^{15}\!+\!x^2\!+\!1$.  

\begin{figure}%[ht]
    \centering
    \includegraphics[width=1\columnwidth]{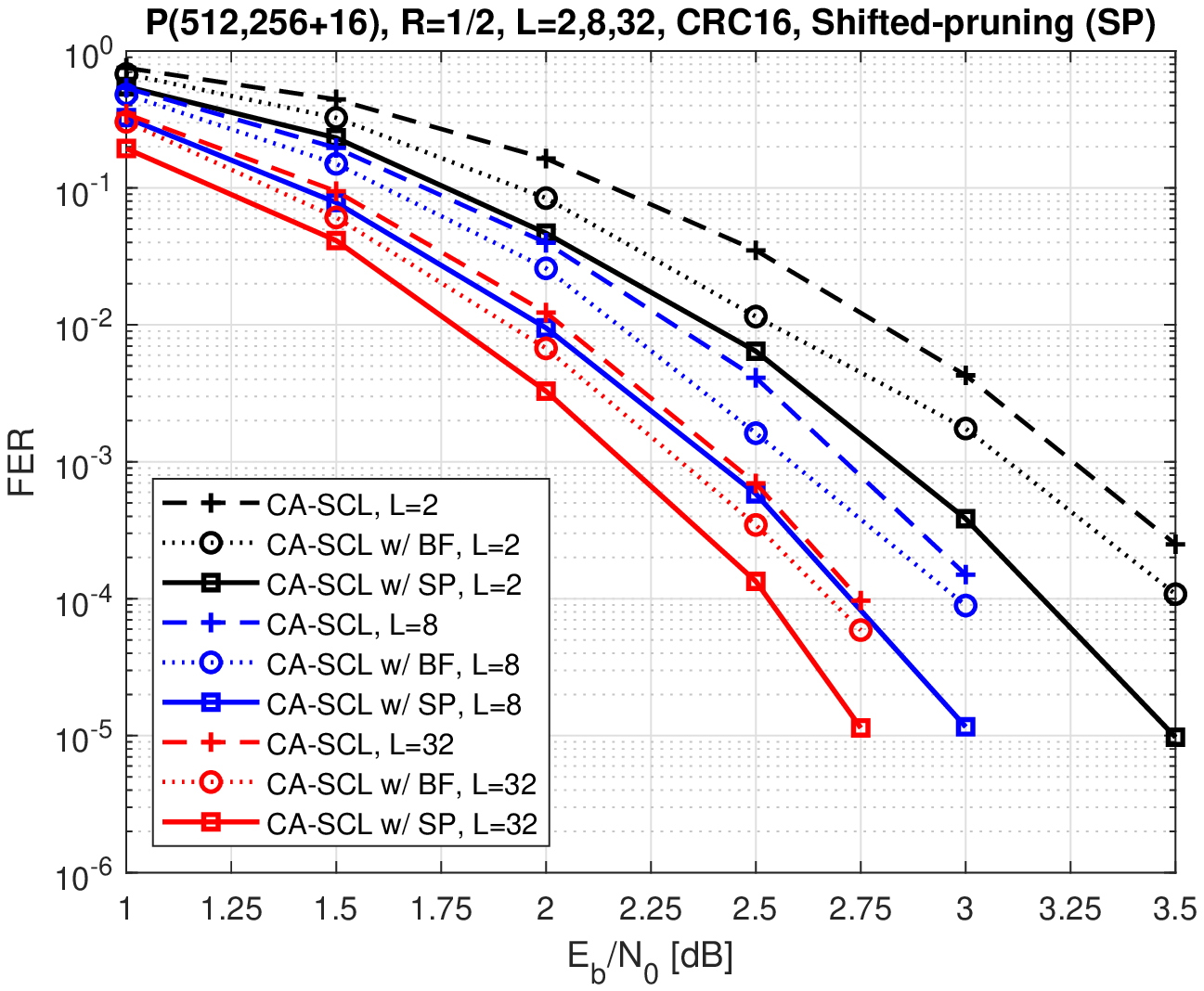}
    %\vspace{-5pt}
    \caption{Error correction performance under conventional CA-SCL decoding, CA-SCL decoding with shifted pruning (SP) and with bit-flipping (BF)}
    \label{fig:perf1}
    \vspace{-5pt}
\end{figure}

\begin{figure}
    %\vspace{10pt}
    \centering
    \includegraphics[width=1\columnwidth]{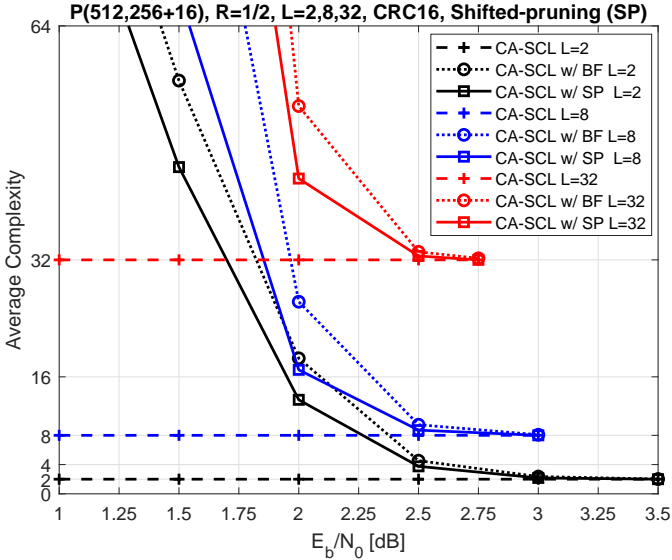}
    %\vspace{-5pt}
    \caption{Average complexity (equivalent to average list size considering additional attempts) under conventional CA-SCL decoding, CA-SCL decoding with shifted pruning (SP) and with bit-flipping (BF)}
    \label{fig:complx}
    \vspace{-5pt}
\end{figure}

Fig. \ref{fig:perf1} compares the error correction performance of $P(512,256+16)$ under conventional CA-SCL decoding and CA-SCL decoding with shifted-pruning (SP) using different list sizes ($L=2,8,32$). The size of the critical sets, $|\mathcal{CS}|$ for code rates 0.5 and 0.8 are 74 and 58 bits, respectively. The maximum number of attempts ($T$) is set to the size of critical set, i.e. $T=|\mathcal{CS}|$. Note that the complexity obtained here is from unsorted $\mathcal{CS}$, i.e. the probability of elimination of the correct path is not used to prioritize the bit indices.  The results in Fig. \ref{fig:perf1} show the gains of about 0.5 dB, 0.35 dB and 0.3 dB at $FER=10^{-4}$ when list size $L$ is 2, 8 and 32, respectively. For comparison, the performance of CA-SCL decoding with bit-flipping (BF) technique \cite{yongrun} for $L=2$ and 8 and $T=|\mathcal{CS}|$ are also shown in Fig. \ref{fig:perf1}.

\begin{figure}%[ht]
    \centering
    \includegraphics[width=1\columnwidth]{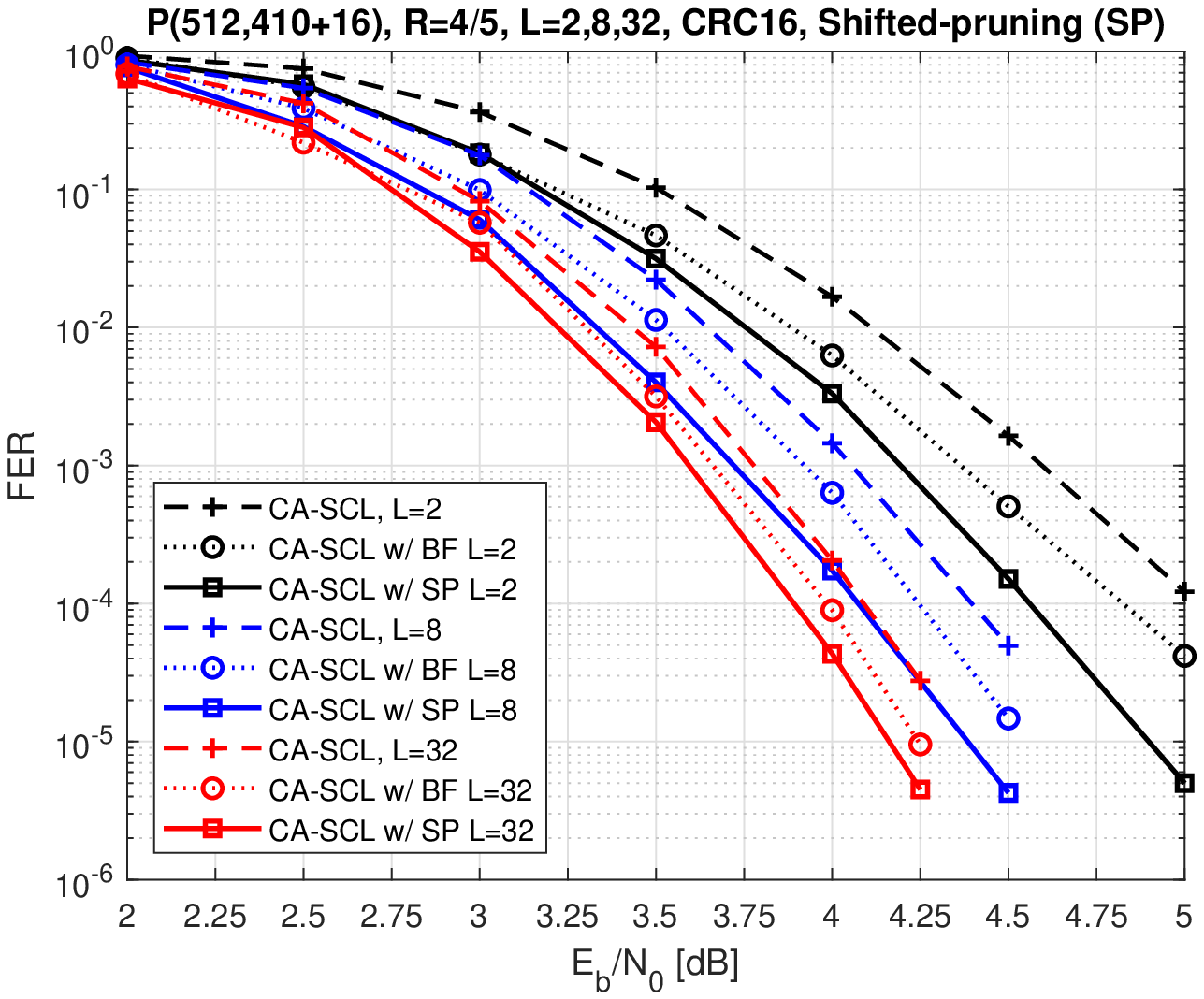}
    \caption{Error correction performance under conventional CA-SCL decoding, CA-SCL decoding with shifted pruning (SP) and with bit-flipping (BF)}
    \label{fig:perf2}
    \vspace{-5pt}
\end{figure}

Fig. \ref{fig:complx} illustrates the average computational complexity for obtaining the curves shown in Fig. \ref{fig:perf1}. Although the average complexity at low $E_b/N_0$ range is significantly high, the complexity at practical $E_b/N_0$ range which provides the error correction performance of $FER<10^{-3}$ is low. For example, for $L=8$ (the blue curves in Fig.\ref{fig:perf1}), the performance is practically acceptable at $E_b/N_0\geq 2$ dB. As Fig. \ref{fig:perf2} shows, the average complexity at $E_b/N_0\geq 2$ dB is close or almost equal to the complexity of conventional SCL decoding.

Similar gains in the error correction performance are obtained for $P(512,410+16)$ as shown in Fig. \ref{fig:perf2}.

\begin{figure}%[ht]
    \centering
    \includegraphics[width=1\columnwidth]{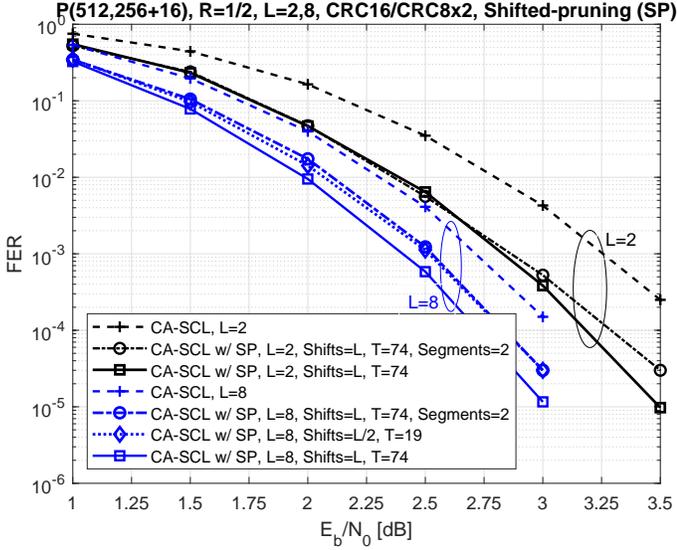}
    \caption{Error correction performance under segmented CA-SCL decoding with shifted pruning (SP) and CA-SCL decoding with constrained shifted pruning (SP)}
    \label{fig:perf3}
   % \vspace{-10pt}
\end{figure}

Aiming to reduce the computational complexity of CA-SCL decoding under shifted-pruning, we implement the segmented CA-SCL decoder hoping to realize the nested shifted-pruning scheme. Fig. \ref{fig:perf3} shows the performance of this implementation for $P(512,256+8x2)$ using two segments protected by two 8-bit CRCs with  generator polynomial $g(x)=x^{8}\!+\!x^{7}\!+\!x^6+\!x^4+\!x^2\!+\!1$. The code was constructed by DE/GA method and design-SNR=4.

\begin{figure}%[ht]
    \centering
    \includegraphics[width=1\columnwidth]{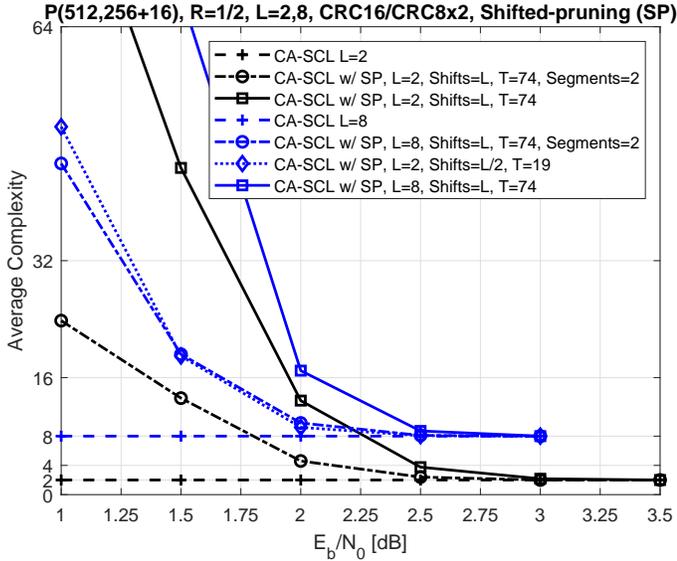}
    \caption{Average complexity  under segmented CA-SCL decoding with shifted pruning (SP) and CA-SCL decoding with constrained shifted pruning (SP)}
    \label{fig:complx2}
    %\vspace{-10pt}
\end{figure}

As can be seen, the performance for list size $L=2$ overlaps with the non-segmented decoding except at high SNRs, while there is a gap of about 0.1 dB for $L=8$. This gap appears due to high probability of undetected error when using short CRCs, particularly when the number of candidates to check at large list size is significantly more than $L=2$. On the other hand, this slight degradation  can be traded with a significant reduction in the computational complexity as shown in Fig. \ref{fig:complx2}.

According to section \ref{sec:generalized SP}, by careful choosing of a limited number of bit-positions for shifting the pruning window, we can reduce the computational complexity at a small cost of performance degradation. The results shown in Fig. \ref{fig:perf3} and \ref{fig:complx2} for shifting the pruning window by $L/2$ positions, instead of $L$ positions in the previous results, over 19 bit-positions illustrate the expected reduction in the complexity similar to segmented CA-SCL decoding at a similar degradation cost. 

\section{Conclusion}
In this paper, we analyze the elimination of the correct path in the list decoding process. Then, the shifted-pruning scheme is proposed to significantly reduce the probability of elimination of the correct path in additional decoding attempts when the decoding fails. The numerical results for polar codes of length $N=2^9$ show that at $FER<10^{-2}$, the proposed scheme with $L$-position shift offers a significant gain of 0.25 dB to 0.5 dB depending on the list size and the code rate, while the average computational complexity remains close to the computational complexity of the conventional CA-SCL decoding at this FER range. Additionally, a generalized form of shifted-pruning is proposed where the shift is variable, e.g. $L/2$, instead of $L$. The result for a constrained set of bit-positions for performing shifted-pruning scheme shows a significant reduction in the computational complexity at low SNR regime by trading it with a small performance degradation. Similar results can be obtained under segmented CA-SCL decoding without need to constrain the bit-position set. 
%\todo{\tiny Check references: [9] update}

\addtolength{\textheight}{-12cm}   % This command serves to balance the column lengths
                                  % on the last page of the document manually. It shortens
                                  % the textheight of the last page by a suitable amount.
                                  % This command does not take effect until the next page
                                  % so it should come on the page before the last. Make
                                  % sure that you do not shorten the textheight too much.

%%%%%%%%%%%%%%%%%%%%%%%%%%%%%%%%%%%%%%%%%%%%%%%%%%%%%%%%%%%%%%%%%%%%%%%%%%%%%%%%

%%%%%%%%%%%%%%%%%%%%%%%%%%%%%%%%%%%%%%%%%%%%%%%%%%%%%%%%%%%%%%%%%%%%%%%%%%%%%%%%

%%%%%%%%%%%%%%%%%%%%%%%%%%%%%%%%%%%%%%%%%%%%%%%%%%%%%%%%%%%%%%%%%%%%%%%%%%%%%%%%
%\section*{APPENDIX}

%Appendixes should appear before the acknowledgment.

%\section*{ACKNOWLEDGMENT}
%The authors would like to thank Yu Yongrun from Southeast University in China for sharing the MATLAB code of SC list bit-flip decoder.


\begin{thebibliography}{9}

\bibitem{arikan} E. Arikan, ``Channel polarization: A method for constructing capacity-achieving codes for symmetric binary-input memoryless channels," {\em IEEE Trans. Inf. Theory,} vol. 55, no. 7, pp. 3051-3073, Jul. 2009.
%\bibitem{3GPP} ``Final report of 3GPP TSG RAN WG1 \#87 v1.0.0," http://www.3gpp.org/ftp/tsg ran/WG1 RL1/TSGR1 87/Report/Final Minutes report RAN1\%2387 v100.zip, Reno, USA, November 2016.
\bibitem{tal} I. Tal and A. Vardy, ``List decoding of polar codes," {\em IEEE Int. Symp. on  Information Theory,} St. Petersburg, Russia, Jul. 2011, pp. 1–5.
%\bibitem{mori} R. Mori and T. Tanaka, ``Performance and construction of polar codes on symmetric binary-input memoryless channels," in Proc. IEEE ISIT, Jun./Jul. 2009, pp. 1496–1500.
%\bibitem{tal2} I. Tal and A. Vardy, ``How to construct polar codes," IEEE Trans. Inf. Theory, vol. 59, no. 10, pp. 6562–6582, Oct. 2013.
%\bibitem{he} G. He, J. C. Belfiore, X. Liu, Y. Ge, R. Zhang, I. Land, Y. Chen, R. Li, J. Wang, G. Yang, and T. Wen, ``$\beta$-expansion: A theoretical framework for fast and recursive construction of polar codes," in Global Communications Conference (GLOBECOM), 2017 IEEE. IEEE, 2017, pp. 1–7.
%\bibitem{mondelli} M. Mondelli, S. H. Hassani, and R. Urbanke, ''Construction of polar codes with sublinear complexity," in Proc. IEEE Int. Symp. Inform. Theory, Jun. 2017, pp. 1853–1857
%\bibitem{qin} M. Qin, J. Guo, A. Bhatia, A. G. i Fabregas, and P. Siegel, ``Polar code constructions based on LLR evolution," IEEE Communications Letters, vol. PP, no. 99, pp. 1–1, 2017.
\bibitem{afisiadis} O. Afisiadis, A. Balatsoukas-Stimming, and A. Burg, ``A low-complexity improved successive cancellation decoder for polar codes," {\em IEEE 48th Asilomar Conference on Signals, Systems and Computers,} 2014, pp. 2116-2120.
%\bibitem{niu} K. Niu and K. Chen, ``Stack decoding of polar codes," {\em IET Electronics Letters,} vol. 48, no. 12, pp. 695-696, 2012.
%\bibitem{alamdar-yazdi} A. Alamdar-Yazdi and F. R. Kschischang, ``A simplified successive cancellation decoder for polar codes," {em IEEE Commun. Lett.,} vol. 15, no. 12, pp. 1378-1380, Dec. 2011.
%\bibitem{sarkis} G. Sarkis, P. Giard, A. Vardy, Claude Thibeault, and W. J. Gross, ``Fast list decoders for polar codes," {\em IEEE J Sel Areas Commun,} vol. 32, no. 5, pp. 946-957, May 2014.
%\bibitem{k_chen} K. Chen, B. Li, H. Shen, J. Jin, and D. Tse, ``Reduce the complexity of list decoding of polar codes by tree-pruning," {\em IEEE Commun. Lett.,} vol. 20, no. 2, pp. 204-207, Feb. 2016.
%\bibitem{j_chen} J. Chen, Y.Z. Fan, C.Y. Xia, C.Y. Tsui, J. Jin, K. Chen, and B. Li, ``Low-Complexity List Successive-Cancellation Decoding of Polar Codes Using List Pruning," {\em IEEE Global Communications Conference,} Washington DC, USA, Dec. 2016, pp. 1-6.
%\bibitem{yuan} B. Yuan and K. K. Parhi, ``Low-latency successive-cancellation list decoders for polar codes with multibit decision," {\em IEEE Trans. Very Large Scale Integr. Syst.,} vol. 23, no. 10, pp. 2268-2280, Oct. 2015.
%\bibitem{xiong} C. Xiong, J. Lin, and Z. Yan, ``Symbol-decision successive cancellation list decoder for polar codes," {\em IEEE Trans. Signal Process.,} vol. 64, no. 3, pp. 675-687, Feb. 2016.
%\bibitem{zhang} Z. Zhang, L. Zhang, X. Wang, C. Zhong, H. V. Poor, ``A split-reduced successive cancellation list decoder for polar codes," {\em IEEE J. Select. Areas Commun.,} vol. 34, no. 2, pp. 292-302, Feb. 2015.
%\bibitem{guo} J. Guo, Z. Shi, Z. Liu, Z. Zhang, Q. Liu, ``Multi-CRC polar codes and their applications," {\em Commun. Lett.,} vol. 20, no. 2, pp. 212-215, Feb. 2016.
%\bibitem{hashemi} S. A. Hashemi, C. Condo, F. Ercan, and W. J. Gross, ``Memory-efficient polar decoders," {\em  IEEE Journal on Emerging and Selected Topics in Circuits and Systems,} vol. 7, no. 4, pp. 604-615, Dec. 2017.
%\bibitem{hashemi} S. A. Hashemi, A. Balatsoukas-Stimming, P. Giard, C. Thibeault, and W. J. Gross, ``Partitioned Successive-Cancellation List Decoding of Polar Codes," {\em  IEEE Inter. Conf. on Acoustics Speech and Sig. Process. (ICASSP),} Shanghai, China, March 2016, pp. 957-960.
%\bibitem{rowshan} M. Rowshan, E. Viterbo, et. al, ``Logarithmic Non-uniform Quantization for List Decoding of Polar Codes,"  submitted to {\em IEEE Communications Letters}.
%\bibitem{Zhang2} Q. Zhang, A. Liu and X. Pan, ``Efficient CRC concatenation scheme for polar codes," {\em Electronics Letters,} vol. 53, no. 13, pp. 860–862, Jun. 2017.
%\bibitem{leroux} C. Leroux, A. J. Raymond, G. Sarkis, and W. J. Gross, ``A semi-parallel successive-cancellation decoder for polar codes," {\em IEEE Trans. Signal Process.,} vol. 61, no. 2, pp. 289-299, Jan. 2013.
\bibitem{chandesris} L. Chandesris, V. Savin, and D. Declercq, ``An improved SCflip decoder for polar codes,`` in {\em 2016 IEEE Global Communications Conference (GLOBECOM),} Dec. 2016, pp. 1–6.
\bibitem{zhang} Z. Zhang, K. Qin, L. Zhang, H. Zhang, and G. T. Chen, ``Progressive bit-flipping decoding of polar codes over layered critical sets,`` in  {\em IEEE Global Communications Conference (GLOBECOM),} Dec 2017, pp. 1–6.
%\bibitem{ercan} F. Ercan, C. Condo and W. J. Gross, ``Improved Bit-Flipping Algorithm for Successive Cancellation Decoding of Polar Codes," in {\em IEEE Transactions on Communications,} vol. 67, no. 1, pp. 61-72, Jan. 2019.
\bibitem{yongrun} Y. Yongrun, P. Zhiwen, L. Nan and Y. Xiaohu, ``Successive Cancellation List Bit-flip Decoder for Polar Codes,`` {\em 10th International Conference on Wireless Communications and Signal Processing (WCSP),} Hangzhou, 2018, pp. 1-6.
\bibitem{li} B. Li, H. Shen, and D. Tse, ``An adaptive successive cancellation list decoder for polar codes with cyclic redundancy check," {\em IEEE Communications Letters,} vol. 16, no. 12, pp. 2044–2047, Dec. 2012.
\bibitem{rowshan2} M. Rowshan, E. Viterbo, R. Micheloni and A. Marelli, ``Repetition-assisted decoding of polar codes," {\em Electronics Letters,} vol. 55, no. 5, pp. 270-272, 7 3 2019.
\bibitem{balatsoukas} A. Balatsoukas-Stimming, M. Bastani Parizi, and A. Burg, ``LLR-based successive cancellation list decoding of polar codes," {\em IEEE Trans. Signal Processing,} vol. 63, no. 19, pp. 5165-5179, Oct 2015.
\bibitem{rowshan} M. Rowshan and E. Viterbo, ``Stepped List Decoding for Polar Codes," {\em 2018 IEEE 10th International Symposium on Turbo Codes \& Iterative Information Processing (ISTC),} Hong Kong, Hong Kong, 2018, pp. 1-5.
\bibitem{rowshan3} M. Rowshan, E. Viterbo, ``How to Modify Polar Codes for List Decoding," {\em IEEE Int. Symp. on  Information Theory,} Paris, France, Jul. 2019, pp. 1772-1776.
%\bibitem{sarkis} G. Sarkis, P. Giard, A. Vardy, C. Thibeault and W. J. Gross, ``Fast Polar Decoders: Algorithm and Implementation," in {\em IEEE Journal on Selected Areas in Communications}, vol. 32, no. 5, pp. 946-957, May 2014.
%\bibitem{kahraman} S. Kahraman, E. Viterbo and M. E. Çelebi, ``Folded tree maximum-likelihood decoder for Kronecker product-based codes," {\em 2013 51st Annual Allerton Conference on Communication, Control, and Computing (Allerton),} Monticello, IL, 2013, pp. 629-636.
\bibitem{guo} J. Guo, Z. Shi, Z. Liu, Z. Zhang, Q. Liu, ``Multi-CRC polar codes and their applications," {\em Commun. Lett.,} vol. 20, no. 2, pp. 212-215, Feb. 2016.
\bibitem{trifonov} P. Trifonov, ``Efficient design and decoding of polar codes," {\em IEEE Trans. Commun.,} vol. 60, no. 11, pp. 3221–3227, Nov. 2012.

\end{thebibliography}
\end{document}